\documentclass{jfm}
\usepackage{color}

\begin{document}

\newtheorem{lemma}{Lemma}
\newtheorem{corollary}{Corollary}

\shorttitle{The Potential Energy Density of Multicomponent Compressible Stratified Fluids} 
\shortauthor{R. Tailleux} 

\title{Local available energetics of multicomponent compressible stratified fluids}

\author
 {
R\'emi Tailleux \aff{1}
  \corresp{\email{R.G.J.Tailleux@reading.ac.uk}},
  }

\affiliation
{
\aff{1}
Dept of Meteorology, University of Reading, Earley Gate, PO Box 243, Reading RG6 6BB, UK
}

\maketitle

\begin{abstract}
\textcolor{blue}{We extend} the local theory of available potential energy (APE)
to a general multicomponent compressible stratified fluid,
accounting for the effects of diabatic sinks and sources.
As for simple compressible fluids, the total potential energy density \textcolor{blue}{of a} fluid parcel is the sum of its
available elastic energy \textcolor{blue}{(AEE)} and APE density. 
\textcolor{blue}{These} respectively represent the adiabatic compression/expansion
work needed to bring it from its reference pressure to its actual pressure and the work against buoyancy forces 
required to move it from its reference state position to its actual position. 
\textcolor{blue}{Our expression for the APE density is new and 
derived} using only elementary manipulations of the equations of motion; \textcolor{blue}{it}
is significantly simpler than existing published expressions, 
while also being more transparently linked to the relevant form of APE density
for the Boussinesq and hydrostatic primitive equations. \textcolor{blue}{Our} new framework is used to clarify the links between some
aspects of the energetics of Boussinesq and real fluids, \textcolor{blue}{as well as to shed light on the physical basis underlying the
choice of reference state(s) in local APE theory.} 
%
%
%
\end{abstract}

\vspace{-0.8cm}

\section{Introduction}

The concept of available potential energy (APE), first developed by \citet{Lorenz1955} and \citet{Margules1903},
aims to quantify the part of the total potential energy (PE) of a stratified fluid that is available for (reversible) conversions
with kinetic energy (KE). In Lorenz's view, any realisable 
state of a stratified fluid is viewed as an adiabatic re-arrangement of some notional rest state, whose background
potential energy $PE_r$ is `inert' or `unavailable' in some sense, thus motivating the
partition $PE=APE+PE_r$. From a fundamental viewpoint, the concept of APE provides a natural solution to 
a number of outstanding problems plaguing the standard PE. 
Indeed, in contrast to the latter: 1) APE is independent of the arbitrary reference level defining geopotential height and of the 
unknowable constants generally entering the construction of internal energy; 
2) APE is naturally positive definite and at least quadratic in perturbations, thus allowing it to be meaningfully partitioned 
into mean and eddy components;  3) APE does not suffer from the `cooling' paradox, \textcolor{blue}{which occurs 
when} a given forcing acts as a sink of potential energy (such as high-latitude cooling in the oceans), 
but nevertheless results in the creation of kinetic energy owing to such a forcing being a source of APE; 
4) the APE budget naturally yields a prediction for the
thermodynamic efficiency of the system considered that appears to be physically much more meaningful than that 
predicted by the entropy budget, as discussed by \cite{Tailleux2010}. This is particularly important for understanding how
to define and quantify the power input due to surface buoyancy fluxes in the ocean for instance.
It is no surprise, therefore, that Lorenz APE theory has been and still primarily remains the main tool for discussing the
energy cycle of the atmosphere and oceans, e.g., \cite{Peixoto1992}

A longstanding difficulty with Lorenz APE theory, however, is that it is only global in nature. This is an important drawback
\textcolor{blue}{complicating} its use in local or regional studies of energetics, which has historically prompted 
much research effort into identifying a suitable local extension,
the main attempts being reviewed in \citet{Tailleux2013}. \textcolor{blue}{The most successful attempts, and most clearly
related to Lorenz APE theory,} are the local frameworks proposed by \citet{Andrews1981} for a simple compressible
stratified fluid and by \citet{Holliday1981} for a Boussinesq fluid with a linear equation of state; \textcolor{blue}{these 
attempts, along with other related older formulations, were subsequently unified within the context of Hamiltonian theory by \citet{Shepherd1993}}.
Yet, despite having been established over 35 years ago, it is only relatively recently that local theories of APE have
started to receive attention in the context of stratified turbulence \citep{Roullet2009,Molemaker2010,
Scotti2014,Winters2013}, ocean energetics \citep{Scotti2006,Tailleux2013b,Roullet2014,Saenz2015,Zemskova2015,MacCready2016}, 
and atmospheric energetics \citep{Kucharski1997,Kucharski2000,Peng2015,Novak2018}. 

Despite all recent advances and increased interest, however, a satisfactory generalisation of 
\citet{Andrews1981}'s local APE theory to a multicomponent compressible stratified fluid accounting for
diabatic sources and sinks \textcolor{blue}{has remained out of reach.} Yet, such a generalisation is essential for the
development of any rigorous
theoretical understanding of the role played by salinity or humidity, as well as of turbulent mixing, in the oceanic and atmospheric 
energy cycles. \citet{Bannon2003} attempted such a generalisation by means of \citet{Shepherd1993}'s Hamiltonian framework,
but \textcolor{blue}{his} theory only \textcolor{blue}{pertains to an ideal fluid whose constituents are all independent of horizontal
position in the reference state, thus excluding the possibility of density compensation \citep{Tailleux2005}}, 
a key aspect of real fluids.
Recently, \citet{Peng2015} have proposed a local 
APE theory for a moist atmosphere, but it is not exact and limited to the anelastic approximation. 
How to achieve such a generalisation by means of elementary manipulation of the full Navier-Stokes equations
is the main aim of the present paper, and is derived in Section 2. 
Section 3 discusses the effects of diabatic sources
and sinks and the links between the energy conversions taking place in real and Boussinesq fluids.
\textcolor{blue}{Section \ref{reference_states} reviews and discusses the physical issues underlying the construction of the reference
state(s) in local APE theories.}
Section \ref{discussion} \textcolor{blue}{summarises and discusses} the results.

\vspace{-0.5cm}

\section{Available energetics of a stratified compressible binary fluid}
\label{theory}

\textcolor{blue}{We start by revisiting \citet{Andrews1981}'s construction of potential energy (PE) density so as to 
generalise it easily to multi-component compressible stratified fluids. }
As in \citet{Andrews1981} or \citet{Bannon2003}, we define
the PE density of a stratified fluid as the sum of its Available Elastic Energy (AEE) density and APE density.
As showed below, this can be achieved without 
loss of generality by restricting attention to the case of a binary fluid, that is, one whose equation of state
for density depends on \textcolor{blue}{composition} \textcolor{blue}{$S$}
(referred to as `salinity' \textcolor{blue}{hereafter}) in addition to specific
entropy \textcolor{blue}{$\eta$} and pressure \textcolor{blue}{$p$, all quantities being functions of spatial position
$(x,y,z)$ and time $t$}. A full set of governing equations in
Earth rotating frame is:
\begin{equation}
      \rho \frac{D{\bf v}}{Dt} + 2{\bf \Omega} \times \rho {\bf v}
       + \nabla p = - \rho \nabla \Phi + \rho {\bf F} ,
      \label{momentum}
\end{equation}
\begin{equation}
   \nabla \cdot {\bf v} = \frac{1}{\upsilon}\frac{D\upsilon}{Dt} ,
   \label{continuity}
\end{equation}
\begin{equation}
     \frac{D(\eta,S)}{Dt} = (\dot{\eta},\dot{S}) , 
     \label{thermodynamics}
\end{equation}
\begin{equation}
     \upsilon = \upsilon(\eta,S,p) .
     \label{eos}
\end{equation}
Here, ${\bf v} = (u,v,w)$ is the velocity field, ${\bf \Omega}$ is Earth rotation vector,
$\Phi(z) = g_0 z$ is the geopotential ($g_0$ is gravitational
acceleration), $z$ is height increasing positively upwards, $\upsilon = 1/\rho$ is the
specific volume, $\rho$ is density, ${\bf F}$ is frictional force, while $\dot{\eta}$ and $\dot{S}$ represent diabatic
sources of entropy and salt. 

We first address the case of an ideal fluid, thus
setting ${\bf F}=\dot{\eta}=\dot{S} = 0$ in (\ref{momentum}-\ref{eos}). 
\textcolor{blue}{In this case}, a standard expression for the total energy budget is:
\begin{equation}
        \rho \frac{D}{Dt} \left ( \frac{{\bf v}^2}{2} + \Phi + e \right ) + \nabla \cdot ( p {\bf v} )  = 0 ,
        \label{standard_energy_budget} 
\end{equation}
where $e = e(\eta,S,p) = h(\eta,S,p) - p \upsilon$ is the specific internal energy and $h$ the specific enthalpy.
In order to obtain an expression for the PE density of the fluid considered, one first needs to identify \textcolor{blue}{
suitably defined reference specific volume} $\upsilon_0(z)$ and pressure $p_0(z)$ profiles 
functions of $z$ only and in mechanical (hydrostatic) equilibrium:
 \begin{equation}
      \upsilon_0(z) \frac{\partial p_0}{\partial z}(z) = - \frac{\partial \Phi}{\partial z}(z) .
      \label{reference_profiles}
\end{equation}
\textcolor{blue}{A discussion of the physical issues pertaining to the construction of the reference state is deferred
to Section \ref{reference_states}.} 
The next key step is to assign to each fluid parcel a reference position $z_r(\eta,S)$ in that reference state.
This can be done most generally by regarding $z_r$ as the particular depth at
which the specific volume of the fluid parcel moved adiabatically to that level matches that of the
reference state. Mathematically, this is equivalent to \textcolor{blue}{defining} $z_r$ as a solution of the 
so-called Level of Neutral Buoyancy (LNB) equation
\begin{equation}
     \upsilon ( \eta, S, p_r ) = \upsilon_0(z_r) ,
     \label{LNB_equation}
\end{equation}
similarly as in \citet{Tailleux2013b}, where $p_r(\eta,S) = p_0(z_r(\eta,S))$. \textcolor{blue}{Depending on
how the reference specific volume $\upsilon_0(z)$ is defined}, 
Eq. (\ref{LNB_equation}) \textcolor{blue}{may have no solution, one solution, or even multiple solutions.
In the former case, $z_r$ 
should be defined either at the top or bottom of the domain depending on the buoyancy of the fluid parcel
in the actual state \citep{Tailleux2013b}. The latter case is rarely encountered in the ocean
\citep{Saenz2015} but is commonplace in a moist atmosphere \citep{Wong2016,Harris2018}. How to
select the solution in this case is still poorly understood, however.} 
\par
Following \citet{Andrews1981}, we next use the following identity 
\begin{equation}
    \nabla \cdot \left ( p_0 {\bf v} \right ) = {\bf v} \cdot \nabla p_0 - \frac{p_0}{\rho}\frac{D\rho}{Dt}
     = \rho \frac{D}{Dt} \left ( \frac{p_0}{\rho} \right ) ,
\end{equation}
to rewrite the standard energy budget (\ref{standard_energy_budget}) in terms of the perturbation 
pressure $p-p_0$ as follows:
\textcolor{blue}{
\begin{equation}
   \rho \frac{D}{Dt} \left ( \frac{{\bf v}^2}{2} + {\cal B} \right ) + \nabla \cdot [ (p-p_0) {\bf v} ] = 0,
\end{equation}
where the quantity
\begin{equation}
   {\cal B}(z,S,\eta,p) = \Phi(z) + e(\eta,S,p) + \frac{p_0(z)}{\rho(\eta,S,p)} 
\end{equation}
} is a \textcolor{blue}{hybrid} function of thermodynamic coordinates
and height, reminiscent of the non-kinetic energy part of the classical Bernoulli head. The introduction of ${\cal B}$ proves 
determinant, for it turns out that the quantity
\begin{equation}
   \Pi = \textcolor{blue}{{\cal B} - {\cal B}(z_r,S,\eta,p_r)} 
   = \Phi(z) - \Phi(z_r) + e(\eta,S,p) - e(\eta,S,p_r) + \textcolor{blue}{\frac{p_0(z)}{\rho}} - \frac{p_r}{\rho_r} ,
\end{equation}
obtained as the difference between ${\cal B}$ and
its value in the reference state naturally happens to be positive definite and our sought-for PE density.
The quantity $\Pi + {\bf v}^2/2$ is then \citet{Shepherd1993}'s `pseudo-energy', and obeys the following conservation law:
\begin{equation}
     \rho \frac{D}{Dt} \left ( \frac{{\bf v}^2}{2} + \Pi \right ) 
     + \nabla \cdot [ (p-p_0){\bf v} ] = 0 .
\end{equation}
To prove its positive definite character, it is useful to rewrite $\Pi$ in terms of the
enthalpies $h(\eta,S,p) = e(\eta,S,p) + p/\rho$ and $h(\eta,S,p_r) = e(\eta,S,p_r) + p_r/\rho_r$ as follows:
\begin{equation}
   \Pi = \Phi(z) - \Phi(z_r) + h(\eta,S,p) - h(\eta,S,p_r) + \textcolor{blue}{\frac{p_0(z)-p}{\rho}} ,
\end{equation}
as this naturally yields the decomposition $\Pi = \Pi_1 + \Pi_2$ where
\begin{equation}
     \Pi_1 = h(\eta,S,p) - \textcolor{blue}{h(\eta,S,p_0(z)) + \frac{p_0(z)-p}{\rho}} , 
     \label{pi1_expression}
\end{equation}
\begin{equation}
     \Pi_2 = \Phi(z) - \Phi(z_r) - h(\eta,S,p_r) + \textcolor{blue}{h(\eta,S,p_0(z))} . 
     \label{pi2_expression}
\end{equation}
The quantity $\Pi_1$ is the so-called Available Elastic Energy (AEE) density previously derived by
\citet{Andrews1981} or \citet{Bannon2003}, and may be rewritten as follows:
\begin{equation}
   \Pi_1 = \textcolor{blue}{\int_{p_0(z)}^p} \left [ \upsilon(\eta,S,p') - \upsilon(\eta,S,p) \right ] {\rm d}p'
   = \textcolor{blue}{\int_{p_0(z)}^p} \int_{p'}^p \frac{1}{\rho^2 c_s^2} (\eta,S,p'')\,{\rm d}p'' {\rm d}p' ,
\end{equation}
where the identities ${\rm d}h = T {\rm d}\eta + \mu {\rm d}S + \upsilon {\rm d}p$ and $\upsilon_p = 
- (\rho^2 c_s^2)^{-1}$ have been used. The result $\Pi_1\ge 0$ follows from the $\rho$ and $c_s$ being both
strictly positive quantities. For small pressure 
perturbation $p' = p-p_0$, $\Pi_1$ reduces to the well known quadratic approximation:
\begin{equation}
      \Pi_1 \approx \textcolor{blue}{\frac{(p-p_0(z))^2}{2 \rho^2 c_s^2}} 
\end{equation}
as noted by \citet{Andrews1981} and \citet{Shepherd1993} among others. Physically, $\Pi_1$ represents
the work required to bring the reference pressure $p_0(z)$ of a fluid parcel to its actual pressure $p$ by
means of an adiabatic compression $(p>p_0)$ or expansion $(p<p_0)$. \citet{Andrews1981} argues that
such a term vanishes in the incompressible limit $c_s \rightarrow +\infty$, \textcolor{blue}{thus justifying} its absence in
Boussinesq fluids.

The quantity $\Pi_2$ is the APE density of the fluid and the only part of the PE density generally retained
in incompressible Boussinesq fluids. Physically, it is equivalent to \citet{Andrews1981}'s $\Pi_2$, but 
mathematically, Eq. (\ref{pi2_expression}) is actually much simpler. Moreover, in contrast to \citet{Andrews1981} or
\citet{Bannon2003}'s constructions, it does not require the introduction of 
purely depth-dependent reference profiles for specific entropy and salinity, which is \textcolor{blue}{overly
restrictive in real fluids}.
In order to clarify the link between the APE density of a compressible fluid with that of the Boussinesq or 
hydrostatic primitive equations, we make use of (\ref{reference_profiles}) and of the change of variables 
$p'=p_0(z')$, ${\rm d}p' = -\textcolor{blue}{\rho_0(z')} g_0 {\rm d}z'$, to rewrite $\Pi_2$ in the following mathematically 
equivalent ways: 
$$
    \Pi_2 = - \int_{z_r}^z \upsilon_0(z') \frac{dp_0}{dz}(z') \,{\rm d}z'
        + \textcolor{blue}{\int_{p_0(z_r)}^{p_0(z)}} \upsilon(\eta,S,p') \,{\rm d}p' 
$$
\begin{equation}
    = \textcolor{blue}{\int_{p_0(z_r)}^{p_0(z)}} \left [ \upsilon(\eta,S,p') - \hat{\upsilon}_0(p') \right ]\,{\rm d}p'
   = g_0 \int_{z_r}^z \frac{\left [ \rho(\eta,S,p_0(z')) - \rho_0(z')\right ]}{\rho(\eta,S,p_0(z'))} \,{\rm d}z' ,
  \label{Pi2_equation}
\end{equation}
where $\hat{\upsilon}_0(p) = \upsilon_0(Z_0(p))$, $Z_0(p)$ being the inverse function of $p_0(z)$
satisfying $p_0(Z_0(p)) = p$. Eq. (\ref{Pi2_equation}) are classical expressions for the work against buoyancy
forces required to move a fluid parcel from its reference position at pressure
$p_r = p_0(z_r)$ to its actual position at pressure $p_0(z)$ by means of an adiabatic and isohaline process. 
It is easily seen that the APE density for a Boussinesq fluid derived by \citet{Holliday1981} or 
\citet{Tailleux2013c} can be recovered: 1) by replacing the denominator $\rho(\eta,S,p)$
in the last term of (\ref{Pi2_equation}) by the constant reference Boussinesq density $\rho_{00}$; 2)
by replacing everywhere the reference pressure $p_0(z)$ by the Boussinesq pressure $p_{00}(z) = -\rho_{00} g_0 z$. 
As to the APE density for a hydrostatic dry atmosphere discussed by \citet{Novak2018}, it is simply recovered from
the first term in (\ref{Pi2_equation}) by replacing the upper bound $p_0$ by the hydrostatic pressure $p$ itself.
As a result, \citet{Tailleux2013b}'s arguments may be used to prove the positive definite character
of $\Pi$, as well as its small amplitude approximation $\Pi_2 \approx N_r^2 (z-z_r)^2/2$, with $N_r^2$ given by
\begin{equation}
    N_r^2  = -\frac{g_0}{\rho_r} \left [ \frac{d\rho_0}{dz}(z_r) + \frac{\rho_r g_0}{c_s^2(S,\eta,p_r)} \right ] .
    \label{reference_n2}
 \end{equation}
 Note here that that in a binary or multi-component fluid, the possibility of density compensation means that
 only the reference density and pressure profiles may be assumed to be functions of $z$ alone; all other 
 quantities, including $N_r^2$, must in general depend on horizontal position as well.
\textcolor{blue}{To conclude, let us remark that $\Pi$ is `local' only in the sense of being definable at
any location, since it possesses some degree of non-locality due to being defined relative to 
a `globally-defined' reference state. For this reason, $\Pi$ is best viewed as a sophisticated form of 
density variance, which --- like other standard statistical quantities such
as `anomalies' --- is also nonlocal in some sense.}

\vspace{-0.8cm}

\section{Energy conversions in presence of diabatic sinks and sources} 

We now turn to the issue of how diabatic sinks and sources of $(\eta,S)$ affect the evolution of the
PE density.
To that end, we re-introduce the diabatic terms $(\dot{\eta},\dot{S})$ in the equations for 
$(\eta,S)$. Because the reference state may be altered by diabatic effects, we assume from now on
that the specific volume and pressure reference profiles $\upsilon_0(z,t)$ and $p_0(z,t)$ also
depend on time \citep{Winters1995}. The LNB equation (\ref{LNB_equation}) thus becomes
$\upsilon[\eta,S,p_0(z_r,t)] = \upsilon_0(z_r,t)$ and now defines $z_r = z_r(\eta,S,t)$ as a 
time-dependent material function of $(\eta,S)$. Evolution equations for $\Pi_1$ and $\Pi_2$
are obtained by taking the \textcolor{blue}{material derivatives} of (\ref{pi1_expression}) and (\ref{pi2_expression}).
Adding an evolution equation for the kinetic energy $E_k = {\bf v}^2/2$ (obtained
in the usual way) yields the following description of energetics:
\begin{equation}
     \rho \frac{DE_k}{Dt} + \nabla \cdot [ (p -p_0){\bf v} ] = - \nabla \cdot (p_0{\bf v}) + 
     \frac{p}{\upsilon} \frac{D\upsilon}{Dt} - \rho 
     \frac{D\Phi}{Dt} + \rho {\bf F}\cdot {\bf v} ,
     \label{kinetic_energy}
\end{equation}
\begin{equation}
   \rho \frac{D\Pi_1}{Dt} = \nabla \cdot (p_0{\bf v}) 
    - \frac{p}{\upsilon} \frac{D\upsilon}{Dt} + \rho \dot{\Pi}_1 
   + \left ( 1 - \frac{\rho}{\rho_h} \right ) \frac{\partial p_0}{\partial t}
    + \frac{\rho_0}{\rho_h} \rho \frac{D\Phi}{Dt} ,
    \label{pi1_equation}
\end{equation}
\begin{equation}
    \rho \frac{D\Pi_2}{Dt} = \rho \frac{D\Phi}{Dt} + \rho \dot{\Pi}_2 
      - \frac{\rho_0}{\rho_h} \rho \frac{D\Phi}{Dt} 
      + \frac{\rho}{\rho_h} \frac{\partial p_0}{\partial t}\textcolor{blue}{(z,t)} - \frac{\rho}{\rho_r} 
      \frac{\partial p_0}{\partial t}(z_r,t) .
      \label{pi2_equation}
\end{equation}
In (\ref{pi1_equation}) and (\ref{pi2_equation}), the 
local diabatic production terms $\dot{\Pi}_1$ and $\dot{\Pi}_2$ are defined by
\begin{equation}
   \dot{\Pi}_1 = (T-T_h) \dot{\eta} + (\mu-\mu_h) \dot{S}, \qquad
    \dot{\Pi}_2 = (T_h-T_r) \dot{\eta} + (\mu_h-\mu_r) \dot{S} ,
    \label{PE_dissipations}
\end{equation}
\textcolor{blue}{where $T$ and $\mu$ denote in-situ temperature and relative chemical potential
respectively. Moreover,} the suffix `h' indicates a variable estimated at the pressure $p_0(z,t)$, e.g.,
$T_h = T(\eta,S,p_0(z,t))$, while the suffix `r' indicates a variable estimated at the pressure
$p_r = p_0(z_r,t)$, e.g., $T_r = T(\eta,S,p_r) = T(\eta,S,p_0(z_r,t))$.
Adding (\ref{pi1_equation}) and (\ref{pi2_equation}) yields the following equation for $\Pi$
\begin{equation}
    \rho \frac{D\Pi}{Dt} = \underbrace{\nabla \cdot ( p_0 {\bf v}) + 
    \rho \frac{D\Phi}{Dt} - \frac{p}{\upsilon}
    \frac{D\upsilon}{Dt}}_{C(E_k,\Pi)} + \rho \dot{\Pi} + \frac{\partial p_0}{\partial t}(z,t)
    - \frac{\rho}{\rho_r} \frac{\partial p_0}{\partial t} (z_r,t) ,
\end{equation}
where the local diabatic production/destruction term $\dot{\Pi}$ is defined by:
\begin{equation}
     \dot{\Pi} = (T-T_r) \dot{\eta} + (\mu - \mu_r ) \dot{S} .
     \label{APE_production}
\end{equation}
Note that because the diabatic terms $\dot{S}$ and $\dot{\eta}$ couple $\Pi$ to the background potential 
energy energy $\textcolor{blue}{{\cal B}_r} = \Phi(z_r) + e(\eta,S,p_r) + p_r/\rho_r$, the pseudo-energy $E_k + \Pi$ is no 
longer a conservative quantity. 

\begin{table}
\begin{center}
 \begin{tabular}{ccc}
 Energy Conversion & `Standard' \textcolor{blue}{view} & `Boussinesq' \textcolor{blue}{view} \\[10pt]
 $C(\Pi_1,E_k)$ & $\frac{p}{\upsilon} \frac{D\upsilon}{Dt} - \nabla \cdot ( p_0 {\bf v}) $ & $\frac{p}{\upsilon} \frac{D\upsilon}{Dt}
 - \rho \frac{D\Phi}{Dt} - \nabla \cdot ( p_0 {\bf v} ) $ \\ [5pt]
 $C(E_k,\Pi_2)$ & $\rho \frac{D\Phi}{Dt}$ & 0  \\ [5pt]
 $C(\Pi_2,\Pi_1)$ &  $ \frac{\rho_0}{\rho_h} \rho \frac{D\Phi}{Dt}$ & $\left ( \frac{\rho_0}{\rho_h} -1
 \right ) \rho \frac{D\Phi}{Dt}$ \\ [10pt]
 \label{table_conversion}
 \end{tabular}
 \caption{\textcolor{blue}{The two different possible approaches to defining the energy conversions between} $E_k$, $\Pi_1$
 and $\Pi_2$ in a turbulent multi-component compressible stratified fluid \textcolor{blue}{discussed in this paper}.} 
 \end{center}
 \end{table}
 
We now discuss the nature of 
 the energy conversions between $E_k$, $\Pi_1$ and $\Pi_2$ 
 implied by (\ref{kinetic_energy}-\ref{pi2_equation}) in order to understand its link with Boussinesq 
 energetics. \textcolor{blue}{As is well known, such an exercise is prone to conceptual difficulties, because there is
 no universally agreed way to define energy conversions, which can often be defined in several plausible 
 mathematically equivalent ways.}  In the 
 present case, for instance, there appears to be two natural such ways --- synthesised in 
 Table 1 --- which we refer to as the `standard' and `Boussinesq' \textcolor{blue}{conversions}.
The standard \textcolor{blue}{conversions are those that naturally follow} from the energy budget written as 
 (\ref{kinetic_energy}-\ref{pi2_equation}), and consists in regarding the work of compression/expansion as a conversion
 between KE and AEE and the term $\rho g_0 w$ as a conversion between KE
 and APE. \textcolor{blue}{In this standard view, `large' conversions are permitted to occur between all the
 different energy reservoirs.} \textcolor{blue}{The Boussinesq view, on the other hand, exploits the tendency for
 the standard energy conversions to compensate each other, thus suggesting to recombine the terms of}
(\ref{pi1_equation}) and (\ref{pi2_equation}) as follows:
\begin{equation}
   \rho \frac{D\Pi_1}{Dt} = 
   \underbrace{\nabla \cdot ( p_0 {\bf v}) + \rho \frac{D\Phi}{Dt} - \frac{p}{\upsilon} \frac{D\upsilon}{Dt}}_{C(E_k,\Pi_1)}
 + \rho \dot{\Pi}_1 
  - \underbrace{\left ( 1 - \frac{\rho_0}{\rho_h}  \right ) \rho \frac{D\Phi}{Dt}}_{C(\Pi_1,\Pi_2)} + {\rm N.L},
  \label{new_pi1}
\end{equation}
\begin{equation}
     \rho \frac{D\Pi_2}{Dt} =  \underbrace{\left ( 1 - \frac{\rho_0}{\rho_h} \right ) \rho \frac{D\Phi}{Dt}}_{C(\Pi_1,\Pi_2)}
      + \rho \dot{\Pi}_2 + {\rm N.L.} ,
      \label{new_pi2}
\end{equation}
where the acronym $N.L.$ denotes the non-local terms proportional to the various $\partial p_0/\partial t$
partial derivatives. \textcolor{blue}{The Boussinesq re-organisation has the key advantage of transforming
$C(\Pi_1,\Pi_2)$ into a density flux {\em anomaly}, without actually introducing any form of average. This makes
it arguably more relevant/useful to the study of turbulent stratified fluids than the standard view. 
Physically,} (\ref{new_pi1}-\ref{new_pi2}) no longer allow for any direct conversion between 
kinetic energy and the APE density $\Pi_2$, \textcolor{blue}{so that conversion between KE and $\Pi_2$ appears
to be} indirect and mediated via $\Pi_1$. \textcolor{blue}{As one referee put it, $\Pi_1$ becomes a `pass-through' energy
reservoir.} \textcolor{blue}{The usefulness of this property becomes evident in the incompressible limit 
$c_s \rightarrow +\infty$. Indeed, in this case, $\Pi_1$ vanishes but the energy conversions
in the right-hand side of (\ref{new_pi1}) do not. As a consequence, (\ref{new_pi1}) becomes a diagnostic equation
imposing the following balance to hold approximately at all times}:
\begin{equation}
        \underbrace{\nabla \cdot ( p_0 {\bf v}) + \rho \frac{D\Phi}{Dt} - \frac{p}{\upsilon} \frac{D\upsilon}{Dt}}_{ C(E_k,\Pi_1)}
 + \rho \dot{\Pi}_1 
  - \underbrace{\left ( 1 - \frac{\rho_0}{\rho_h}  \right ) \rho \frac{D\Phi}{Dt}}_{C(\Pi_1,\Pi_2)} + {\rm N.L} \approx 0 . 
  \label{approximate_balance}
\end{equation}
\textcolor{blue}{If $\rho \dot{\Pi}_1$ and $N.L.$ can be neglected} in (\ref{approximate_balance}), 
the above balance \textcolor{blue}{reduces to}
$C(\Pi_1,\Pi_2) \approx C(E_k,\Pi_1)$, which formally allows one to regard the density flux 
\textcolor{blue}{anomaly} $C(\Pi_1,\Pi_2)$ as
a conversion between KE and $\Pi_2$, precisely as \textcolor{blue}{in} Boussinesq energetics. 
Any kinetic energy converted into $\Pi_1$ is instantaneously converted into $\Pi_2$ and conversely.
\textcolor{blue}{To establish that such a view is valid andt physically meaningful, one presumably would need to prove that such
conversions are primarily achieved by sound waves, as these are the only ones capable of transmitting information at infinite
speed in the limit $c_s\rightarrow +\infty$,} 
but this is a challenging technical issue that we regard to be beyond the scope of this paper.

\vspace{-0.8cm}

\textcolor{blue}{
\section{Issues pertaining to the choice of reference state}  
\label{reference_states}
How to define and construct the APE reference state has been a continuing source of confusion and difficulty in APE
theory since its inception. To date, the issue remains controversial and outstanding. Here, we briefly review and
discuss two well known key difficulties in subsections \ref{non_resting_states} and \ref{arbitrariness} below. 
The first one pertains to the
possible use of `non-resting' reference states. The second one pertains to the apparent arbitrariness of the reference
state in local APE theory and whether it implies the possibility to use reference states other than those originally 
proposed by \citet{Lorenz1955}. }

\vspace{-0.5cm}
\textcolor{blue}{
\subsection{Resting versus non-resting reference states} 
\label{non_resting_states}
The standard form of APE is commonly understood to originate from
 horizontal variations in the density field, and therefore naturally
defined relative to a `resting' reference state function of $z$ only (asides possible temporal dependence). A common
objection, however, is that in presence of rotation, idealised fluid flow configurations such as steady 
and stable zonal flows or vortices may have a large fraction of their APE actually locked up by geostrophic or higher 
order type of balance and therefore seemingly unavailable. Such examples have prompted a few authors to look for
a generalisation of APE accounting for momentum constraints, e.g., \citet{Codoban2003,Andrews2006}. Whether 
such an approach is necessarily the `right' one is not clear, however. Indeed, the alternative and much simpler approach
advocated here considers that the definition of APE should not depend on the presence of rotation and flow stability,
and that the potential energy density $\Pi_e$ defined relative to a non-resting reference state should be regarded as 
a form of `eddy' potential energy in a mean/eddy decomposition $\Pi = \Pi_e + \Pi_m + \delta \Pi$ of the total potential
energy density $\Pi$. Such an approach is easily implemented in the present framework by noting that the construction
of $\Pi$ outlined in Section \ref{theory} applies equally well to reference pressure and specific volume 
$p_m(x,y,z)$ and $\upsilon_m(x,y,z)$ that also depend on horizontal position. As a result, the above components of
the partition of $\Pi$ can be explicitly defined by
\begin{equation}
     \Pi_e = \Phi(z) - \Phi(z_{rm}) + e(\eta,S,p) - e(\eta,S,p_{rm}) + \frac{p_m(x,y,z)}{\rho}
      - \frac{p_{rm}}{\rho_{rm}} ,
      \label{eddy_ape}
\end{equation}
\begin{equation}
     \Pi_m = \Phi(z_{rm}) - \Phi(z_r) + e(\eta,S,p_{rm}) - e(\eta,S,p_r) + 
     \frac{p_0(z_{rm})}{\rho_{rm}} - \frac{p_r}{\rho_r} ,
     \label{mean_ape}
\end{equation}
\begin{equation}
    \delta \Pi = \frac{p_0(z)-p_m(x,y,z)}{\rho} + \frac{p_{m}(x,y,z_{rm})-p_0(z_{rm})}{\rho_{rm}}  .
    \label{interaction_ape}
\end{equation}
In Eqs. (\ref{eddy_ape}-\ref{interaction_ape}), $p_r = p_r(\eta,S)$, $\rho_r=\rho_r(\eta,S)$ and 
$z_r=z_r(\eta,S)$ have the same meaning as in Section 
\ref{theory} and are functions of the materially conserved variables $(\eta,S)$ only, 
whereas $z_{rm}$ is the reference depth solution of the LNB equation for the non-resting state
$\upsilon(\eta,S,p(x,y,z_{rm})) = \upsilon_m(x,y,z_{rm})$, with $p_{rm} = p_m(x,y,z_{rm})$. 
Physically, $\Pi_e$ represents the `eddy' potential energy density of a fluid parcel relative to the 
`non-resting' mean state, whereas $\Pi_m$ represents the potential energy density of the fluid parcel in its 
mean equilibrium position relative to its `resting' equilibrium position.
Note that the above mean/eddy partition of $\Pi$ does not explicitly depend
on the introduction of any particular averaging operator, and is valid for arbitrarily defined 'non-resting'
and 'resting' reference state $(p_m,\upsilon_m)$ and $(p_0,\upsilon_0)$; from a mathematical viewpoint,
it is the counterpart of the mean/eddy partitioning of the kinetic energy as 
$ \frac{{\bf v}^2}{2} = \frac{{\bf v}_m^2}{2} + \frac{{\bf v}_e^2}{2} + {\bf v}_m \cdot {\bf v}_e$, 
which also does not require the introduction of any particular averaging operator. The interaction term
$\delta \Pi$ can a priori takes on both signs, and is the counterpart of the interaction 
${\bf v}_m \cdot {\bf v}_e$. 
}
\par
\textcolor{blue}{
Physically, the expectation is that if $(p_m,\upsilon_m)$ correspond to a steady and stable zonal flow or
vortex, rotation should inhibit conversions between $\Pi_m$ and $\Pi_e$. 
To confirm this, let use derive the evolution equations for ${\bf v}^2/2+\Pi_e$, $\Pi_m$ and 
$\delta \Pi$ in the usual manner:
\begin{equation}
  \rho \frac{D}{Dt} \left ( \frac{{\bf v}^2}{2} + \Pi_e \right ) + \nabla \cdot \left \{
  [ p - p_m(x,y,z) ] {\bf v} \right \} 
  =  - \frac{\rho {\bf u}\cdot \nabla_h p_{rm}}{\rho_{rm}} ,
  \label{kinetic_eddy_ape_evolution}
\end{equation}
\begin{equation}
   \rho \frac{D\Pi_m}{Dt} + \nabla \cdot \left \{ \frac{\rho [ p_{rm} - p_0(z_{rm}) ] {\bf v} }{\rho_{rm}}  
   \right \} = \frac{\rho {\bf u}\cdot \nabla_h p_{rm}}{\rho_{rm}} ,
   \label{mean_ape_evolution}
\end{equation}
\begin{equation}
    \rho \frac{D(\delta \Pi)}{Dt} = \nabla \cdot \left \{ \left [ p_0(z)-p_m(x,y,z) 
    +  \frac{\rho (p_{rm}-p_0(z_{rm}))}{\rho_{rm}} \right ]{\bf v} \right \} ,
    \label{interaction_evolution} 
\end{equation}
where $\nabla_h$ denotes the horizontal part of the gradient operator. It is easily verified that summing 
(\ref{kinetic_eddy_ape_evolution}-\ref{interaction_evolution}) recovers (\ref{standard_energy_budget}).
Eqs. (\ref{kinetic_eddy_ape_evolution}-\ref{mean_ape_evolution}) show that the conversion between 
$\Pi_m$ and $\Pi_e$ is controlled by the product of the horizontal velocity times the horizontal gradient
of $p_{rm}$, which vanishes for a steady geostrophically balance state ${\bf u} \propto f_0^{-1} {\bf k} \times
\nabla_h p_{rm}$, where $f_0$ denotes the local rotation rate, and ${\bf k}$ the unit vector pointing 
upwards. Therefore, as far as we can judge, the inhibiting effects of rotation can be understood in the context
of the mean/eddy decomposition of $\Pi$ (see \citet{Scotti2014} and \citet{Novak2018} for different ways of
performing such a decomposition), and does not therefore seem to 
require a definition of APE accounting for momentum constraints. Obviously, this is a subtle and complicated
issue that will warrant extensive discussion in a separate and forthcoming dedicated paper.
}
\textcolor{blue}{
\subsection{On the arbitrariness of the reference state} 
\label{arbitrariness}
The apparent arbitrariness of the reference state in local APE theory, first noted by \citet{Andrews1981}
and \citet{Shepherd1993}, has
been exploited by a few authors to use reference states easier or more convenient to compute than Lorenz
reference state, such as the horizontally-averaged density field \citep{Tailleux2013b} or an analytic 
profile \citep{Peng2015}. Because the choice of reference state affects the magnitude of the APE reservoir,
as well as the rates at which it is created, stored and destroyed \citep{Wong2016}, the present state of affairs
is obviously unsatisfactory and will remain so until the physical basis for how to specify it is identified. The key
role played by ${\cal B}$ in the construction of $\Pi$ suggests that the reference pressure $p_0(z)$ or $p_0(z,t)$
should be chosen so as to approximate the actual pressure field $p(x,y,z,t)$ as accurately as feasible. If least
square were the relevant principle, one would have a physical basis for arguing that the reference density 
field $\rho_0(z,t)$ should be defined as the horizontally averaged density $\overline{\rho}(z,t)$ rather than in
terms of Lorenz reference state. Such an approach would also ensure that $\Pi$ vanishes 
identically in a state of rest, a desirable property of $\Pi$ that is not necessarily satisfied with \citet{Peng2015}'s
analytically specified reference state. Clarifying this issue once and for all therefore remains a key 
priority for future research.
}

\vspace{-0.5cm}

\section{Conclusions}
\label{discussion}
In this paper, we succeeded in extending \citet{Andrews1981}'s local \textcolor{blue}{concept of potential
energy density $\Pi$ to a} multi-component compressible stratified fluid and in deriving evolution equations for the attendant 
energy cycle accounting for diabatic sinks and sources. In contrast to the previous approaches by
\citet{Bannon2003} and \citet{Andrews1981}, our construction
of APE density stands out by its simplicity and economy, as well as by its considerably more transparent link
to the APE density of the Boussinesq and hydrostatic primitive equations. The key step 
making our approach so simple is the introduction of the quantity
${\cal B}(z,\eta,S,p) = \Phi(z) + e(\eta,S,p) + p_0(z) \upsilon(\eta,S,p)$, which is just the standard potential
energy plus the term $p_0/\rho$. \textcolor{blue}{Indeed, all one has to do is to subtract
the reference value of ${\cal B}$ to obtain
$\Pi = {\cal B}(z,\eta,S,p) - {\cal B}(z_r,\eta,S,p_0(z_r))$, a quantity} that is naturally positive definite, 
in contrast to the integrand of \citet{Lorenz1955} globally defined APE. \textcolor{blue}{Such an approach
can be done for any arbitrary resting or non-resting reference state, which points to new ways to
rigorously partition $\Pi$ into mean and eddy components or to think about the role played by rotation on
occasionally inhibiting a large fraction of the total APE. Nevertheless, the exact physical principle(s)
underlying the specification of the reference state(s) remain an outstanding issue in need of further research.}

As in \citet{Andrews1981} and \citet{Bannon2003}, the total potential energy density
of a fluid parcel
is the sum of its available elastic energy (AEE) and APE density, of which only the latter subsists in 
sound-proof systems of equations such as the Boussinesq or hydrostatic primitive equations. 
\textcolor{blue}{This can be formally justified} by noting that AEE 
vanishes in the limit $c_s \rightarrow +\infty$, as pointed out by \citet{Andrews1981}. However, although
AEE vanishes in \textcolor{blue}{the incompressible} limit, this is not in general the case for the individual terms entering its evolution
equation. As a result, the AEE evolution equation becomes a diagnostic equation in the limit $c_s
\rightarrow + \infty$, \textcolor{blue}{which} turns out to be key for establishing a formal link between the energetics of 
Boussinesq and real fluids. 

The present results are important, for they provide for the time the tools needed to tackle many outstanding
issues in the field, ranging from elucidating the role of salt and humidity in the oceanic and 
atmospheric energy cycles, to the development of new energetically and thermodynamically consistent 
mixing parameterisations for use in numerical oceanic and atmospheric models, 
as we hope to demonstrate in future studies. 


\bibliography{jfm-references}
\bibliographystyle{jfm}

\end{document}